# The lack of asymmetry of the Maxwell centroids, and of ocular dominance, in persons with dyslexia

Albert Le Floch[a,b] Guy Ropars[b]

[a] Laboratoire d'électronique quantique et chiralités, 20, square Marcel-Bouget, 35700 Rennes, France

[b] Laboratoire de physique des lasers, UFR SPM, université de Rennes 1, 35042 Rennes, France

Corresponding author :

A. Le Floch, Laboratoire d'électronique quantique et chiralités, 20 square Marcel Bouget, 35700, Rennes, France
Albert.lefloch@laposte.net; guy.ropars@univ-rennes1.fr

## ABSTRACT

While the existence of an asymmetry between the two Maxwell centroids at the centre of the two foveas recorded using a foveascope, leads to the ocular dominance in good readers, the lack of asymmetry in most of the observers with dyslexia leads to their non-dominance and their difficulties in reading and writing. Indeed, the lack of asymmetry between the two main roads to the brain, i.e. the two optical nerves, leads to perturbations in the brain central connectivity, namely between the two hemispheres inducing too robust interhemispheric visual connections, beyond the critical period of 7-8 years. The symmetrical mirror-connections like b-d (observed for about 60% of children with dyslexia) or the non-symmetrical connections like b-b (35%) induce confusions and duplications in observers with dyslexia but remain erasable thanks to pulsed systems (glasses, lamps, screens. . .), using Hebbian processes in primary cortex synapses. The effect is instantaneous, non-invasive and compensates the lack of asymmetry. These systems help most of the observers with dyslexia to overcome their difficulties. The orthoptists and speech therapists can acquire and improve the mechanism to help the children to catch up, and perhaps be able to use the foveascope for an early diagnosis with young children.

*Alors que l'existence d'une asymétrie entre les centroïdes de Maxwell aux centres des deux fovéas, observée grâce au fovéascope, assure la dominance oculaire chez les normolecteurs, le manque d'asymétrie chez les personnes dyslexiques entraine leur manque de dominance oculaire et leurs difficultés de lecture et d'écriture. En effet, ce manque d'asymétrie entre les autoroutes principales au cerveau que sont les deux nerfs optiques, entraine un désordre dans la connectivité neuronale, notamment entre les deux hémisphères, occasionnant des projections interhémisphériques visuelles trop robustes au-delà de la période critique à 7-8 ans. Les projections-miroirs symétriques du type b-d (60 % des cas) ou non-symétriques du type b-b (35 % des cas) engendrent confusions et dédoublements chez les dyslexiques mais restent cependant maitrisables grâce à des systèmes lumineux pulsés (lunettes, lampes, écrans. . .) utilisant les mécanismes Hebbiens aux synapses du cortex. En effet, ces systèmes permettent un contrôle optique du cortex primaire. L'effet est immédiat et non invasif, et compense le manque d'asymétrie. Ces systèmes aident la plupart des dyslexiques à surmonter leurs difficultés. Les orthoptistes et les orthophonistes doivent pouvoir s'approprier et adapter le mécanisme pour aider les enfants à rattraper leur retard, et peut-être même procéder à un diagnostic précoce avec le fovéascope chez les plus jeunes enfants dès 4-5 ans.*

# INTRODUCTION

Dyslexia is a complex disorder [1] that affects about 10% of the population, or 700 million people worldwide. Dyslexic children have difficulty learning to read and write, regardless of their intellectual ability [2]. Vision plays a crucial role in reading and writing [3]. The definitions of dyslexia are not simple (phonological, magnocellular [4], surface, mixed, etc.), the causes are debated, diagnosis often has to wait until the age of 7-8, and there are many different treatments. However, the large majority of specialists agree that dyslexia has a biological and neuronal basis. The brain, with its hundreds of billions of neurons, million of billions of connections and hundreds of brain areas with different functions, plays a crucial role in dyslexia, which seems to be linked to a disorder in neuronal connectivity and too weak lateralization [5,6].

If we compare the flow of information to the brain, we see that the eyes and the two optic nerves are the main 'highways' to the cortex [7]. Each optic nerve has about 1.2 million fibres [8], compared with only 30,000 for each auditory nerve. Despite the apparent symmetry of the two eyes, when we close one of them, we realise that, because of parallax, the two parts of information are slightly different when we look at any given scene. Intuitively, as the brain is unique, we can guess that it will have to rely preferentially on one of the two eyes, the 'dominant eye', breaking the apparent symmetry. This necessary 'symmetry breaking' will have to be based on a certain asymmetry. This asymmetry has been suspected by many authors, such as Ernst Mach [9] who was aware of the difficulties of young children (under 5-6) and of some dyslexics in differentiating between mirror letters such as b and d. He wrote: 'there must be an asymmetry somewhere to remove the confusion between b and d'. Hubel and Wiesel [10,11] explore an 'initial imbalance' to guide ocular dominance columns in the cortex. All of this takes place on the background of the lateralization of the various cognitive functions in the brain, the origin of which is still unknown. For example, the word formation centre in reading (the 'letter box') is located at the base of the left hemisphere for about 97% of people, as shown by Cohen, Dehaene, Naccache et al. in 2000 [12]. To attempt to locate the asymmetry necessary to achieve the 'symmetry breaking' between the two eyes in normal readers, we must first unambiguously identify the dominant eye using the 'noise-activated afterimage method' that we have developed, and then find the asymmetry that determines it. In the hope of justifying the weaker lateralization in dyslexics [5], we could then ask whether there is a lack of asymmetry in dyslexics, and then investigate the implications of this lack of asymmetry for the dyslexia itself, in particular the existence of extra mirror or duplicated images using the noise-activated afterimage method we have already used to define ocular dominance, and possibly propose compensation for this internal visual crowding.

# NOISE-ACTIVATED AFTERIMAGES AND OCULAR DOMINANCE

Ocular dominance has been observed for centuries, in particular by various so-called visual methods using a more or less distant object or scene (Porta, Miles, hole-in-the-map methods,…). However, the results can depend on the distance of the object and on the parallax for some observers, especially dyslexics, which leads to controversies. To overcome these difficulties, we proposed the 'noise-activated afterimages method'. Here, the eyes remain closed after a fixation period of about 10 seconds. We will use the fact that the neurons of the retina and primary cortex are bistable and non-linear, like lasers [13]. A neuron is either at rest or excited ('spike' or action potential). Noise can force a resting neuron to activate, reactivating the negative afterimage following fixation. Furthermore, the noise, i.e. the 2 to 3% of diffuse light that passes through the closed eyelids, can be modulated by alternately placing the hands on one of the two eyelids, allowing the relative power of each eye in relation to the brain to be seen [14]. The eye that sees the clearer afterimage is more strongly connected and is the dominant eye. We first carried out the tests on two cohorts of 30 students each: 30 normal reader students and 30 dyslexic students according their availabilities. All the normal reader students were able to define their dominant eye. The right-eyed/left-eyed ratio was 2, according to various estimates. On the other hand, for the 30 dyslexic students, while the usual methods gave fluctuating results, the activation by noise led to 27 cases out of 30 cases of ocular non-dominance. Three students in the dyslexic cohort had ocular dominance, but unfortunately it was a 'frustrated dominance', because the eye that should have been dominant had been penalised from birth (severe amblyopia in one case, severe strabismus in another, and facial malformation in the third). To investigate the cause of ocular dominance in normal readers and non-dominance in dyslexics, we are going to explore the topography of the different cones at the fundus of the foveas and in particular the specific distribution of blue cones which we have already used in our investigation of Haidinger's brushes [15] which also appear on the foveas when the polarisation of the light is rotated using a polariser [16]. To avoid Newton chromatic dispersion (the blue image is formed slightly in front of the retina), the evolution of the human eye has led to the absence of blue cones in the centre of the fovea, which determines where visual acuity is at its best. This 'blue cone-free area' (see Fig. 1A) is about 100 to 150 μm in diameter and was first indirectly introduced by G. Wald [17] and then directly observed by Curcio and Hendrickson [18], who succeeded in delicately recovering retinas post-mortem, staining the blue cones with antibodies and observed them under the microscope. Unfortunately they were unable to compare the blue cone-free areas in the same person. However, we need in any case to observe the blue cone-free areas in vivo for each observer. Therefore, we built a foveascope to look for any asymmetry in the distribution of blue cones between the two foveas for each observer.

## EXISTENCE OF MAXWELL CENTROID ASYMMETRY

In order to compare these 'blue cone-free areas', we went back to the observations made by Maxwell in 1856 and often overlooked since, the so-called Maxwell's spots [19]. While moving his eye along the exit of a prism in Newton fundamental experiment discovering the decomposition of white light into the colours of the rainbow [20], Maxwell noticed a dark spot when his eye passed into the only blue zone of the spectrum. At the time, he attributed this spot to yellow pigments in the fovea. Maxwell's spot occupies exactly the same area of the fovea where Haidinger's brushes can be seen. The total absence of blue cones in the centre of the foveas corresponds to a small darker zone for the Maxwell spots. In fact, there is no blue detector in this small central region of the Maxwell spot, which we will call the Maxwell centroid.

The foveascope (described in reference 14) that we built to observe Maxwell's spots has a screen (50 cm x 40 cm) brightly illuminated with white light by a 3000 lumen projector. When looking at this screen placed at about 3 m from an eye looking at it through a double filter that is sometimes green, and sometimes blue, each person sees, with each eye, a figure similar to that shown in Fig. 1B, i.e. a Maxwell spot. Only the central part corresponding to the blue cone-free area is of interest here. When we plotted the contour of this Maxwell centroid for each eye, we had our first surprise. For each normal reader, the two centroids are asymmetric. In fact, for a right eye (Fig. 2A) or left eye (Fig. 2B), the contour is quasi-circular for a right eyed observer and for a left eyed observer respectively, whereas for the other eye the contour is elliptical, usually tilted at +/- 45°. The difference in ellipticity, which defines the asymmetry, is therefore measurable. The difference in ellipticity is of the order of + 0.5 for right-eyed observers and - 0.5 for left-eyed observers.

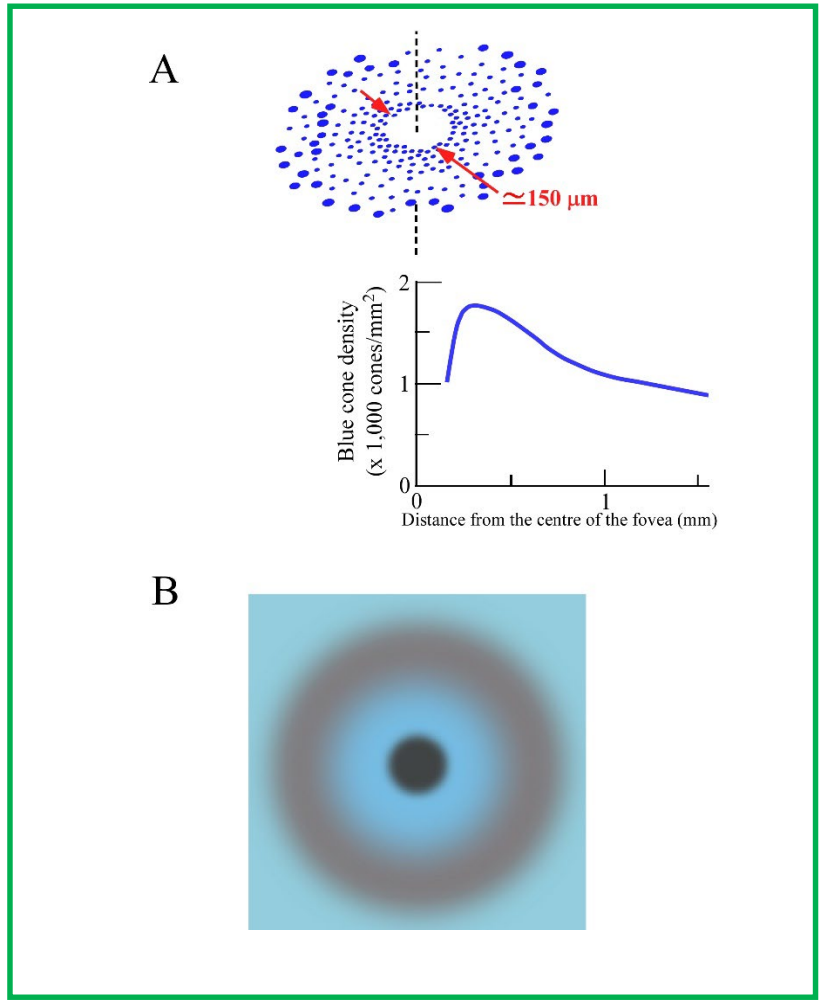

*Figure 1: (A) The 'blue cone-free area' in the distribution of blue cones at the centre of the fovea (taken from [16]). (B) A Maxwell spot, with its centroid, observed with the foveascope.*

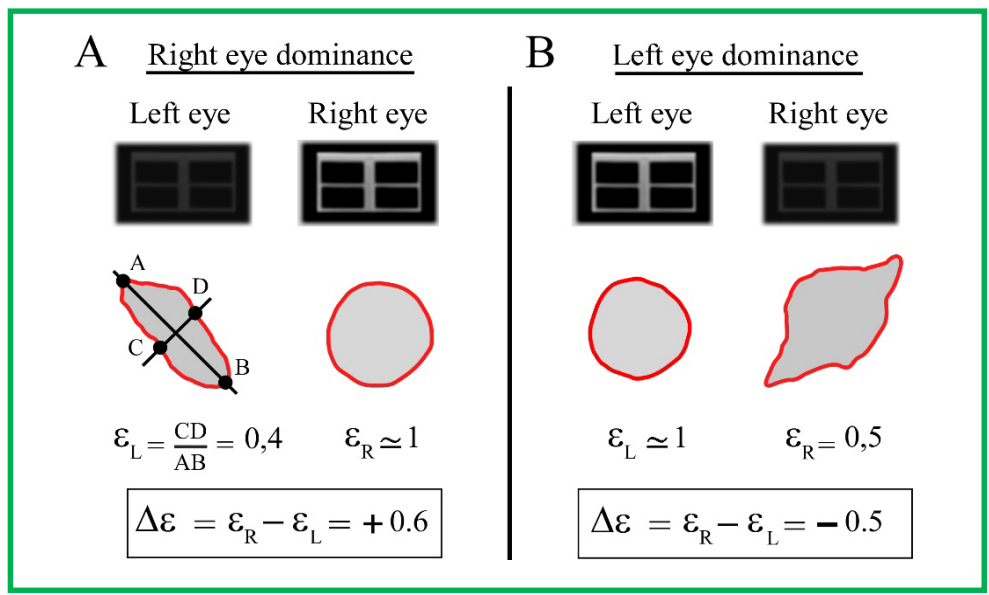

*Figure 2: (A) Afterimages seen by a right-eyed observer and his Maxwell centroids. (B) Afterimages seen by a left-eyed observer and his Maxwell centroids (taken from [14]).*

In addition, the quasi-circular contour always corresponds to the dominant eye determined by the noise-activated afterimage method for normal readers. In contrast, in the cohort of 30 dyslexics, the 27 without ocular dominance had quasi-identical contours for both eyes and no difference in ellipticity (Fig. 3A). In sum, ocular dominance in normal readers is determined by Maxwell centroid asymmetry and the absence of dominance in dyslexics is correlated with the lack of Maxwell centroid asymmetry.

## IMPLICATIONS FOR DYSLEXIA

Disturbed neuronal connectivity has been observed in dyslexics by several authors involving a number of multimodal functions [5]. A comparison of noise-activated afterimages in normal readers and dyslexics will provide crucial information about the nature of the interhemispheric projections needed in humans to connect the two halves of the visual field, and also for understanding most types of dyslexia. If we take the non-symmetrical letter 'b' as a stimulus in Fig. 3B, we can see that for a normal reader, the afterimage activated by the noise is the single letter 'b'. For dyslexics, however, the afterimages include the letter 'b' but also, attenuated, either the letter 'd', or a second duplicated letter 'b' (Fig. 3B). These extra letters 'd' or 'b' correspond to the two types of interhemispheric projections that occur between the two hemispheres of the brain, symmetrical projections giving the 'mirror letter' 'd' (Fig. 3C) and non-symmetrical projections giving the duplicated letter. These projections take place mainly through the corpus callosum, the large bundle of fibres (over 250 million fibres) that connects the two hemispheres. As a result of these projections, b-d confusion persists namely in young children for up to 5-6 years [21]. At the end of the critical period introduced by Hubel and Wiesel [10,11], the asymmetry of the topography of the fovea cones allows the perception of these projections to be erased. However, for dyslexics, the confusion and/or duplication persists beyond the end of the critical period at around 7-8 years of age, resulting in an internal visual crowding which penalises the dyslexic person in terms of reading and writing. In a new cohort of 160 dyslexic children who came to our laboratory for tests and trials, we were able to verify that 60% of the children were disturbed by mirror images and 35% by duplications. Moreover, these projections apply to all letters and even to syllables and words. For example, the afterimage of the word 'neurons' can be accompanied by the noise-activated afterimage 'snoɹuǝu'.

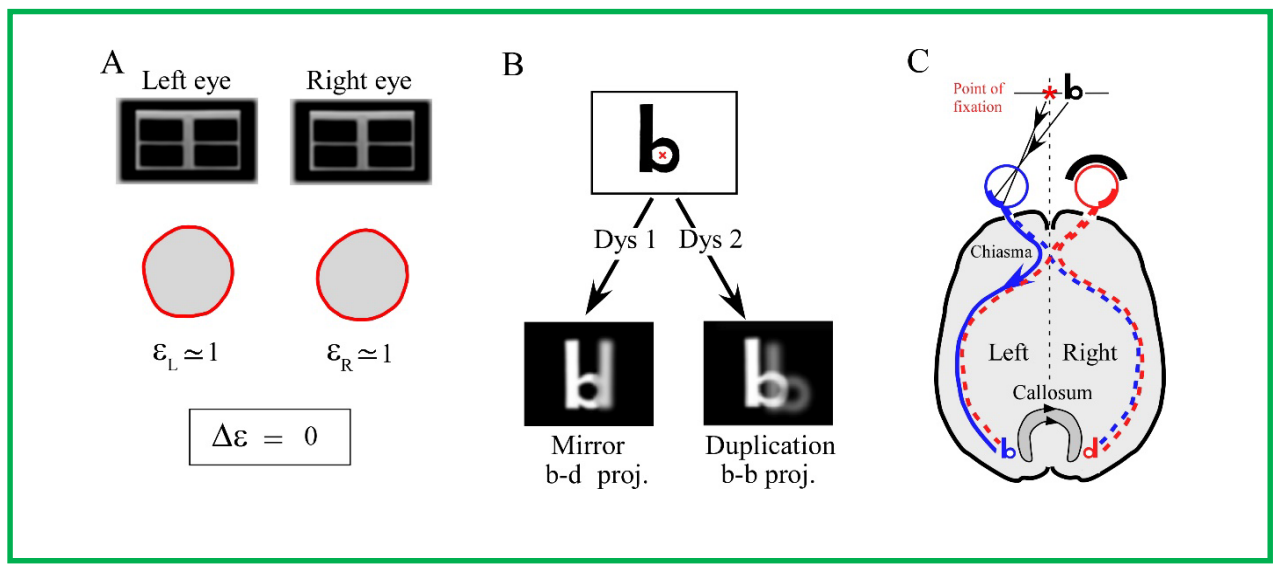

*Figure 3: (A) Afterimages seen by a dyslexic and his Maxwell centroids. (B) Different afterimages of a 'b' seen by two dyslexics (taken from [14]). (C) Example of b-d symmetrical mirror projection through the corpus callosum.*

## COMPENSATION BY OPTICAL CONTROL OF THE PRIMARY CORTEX

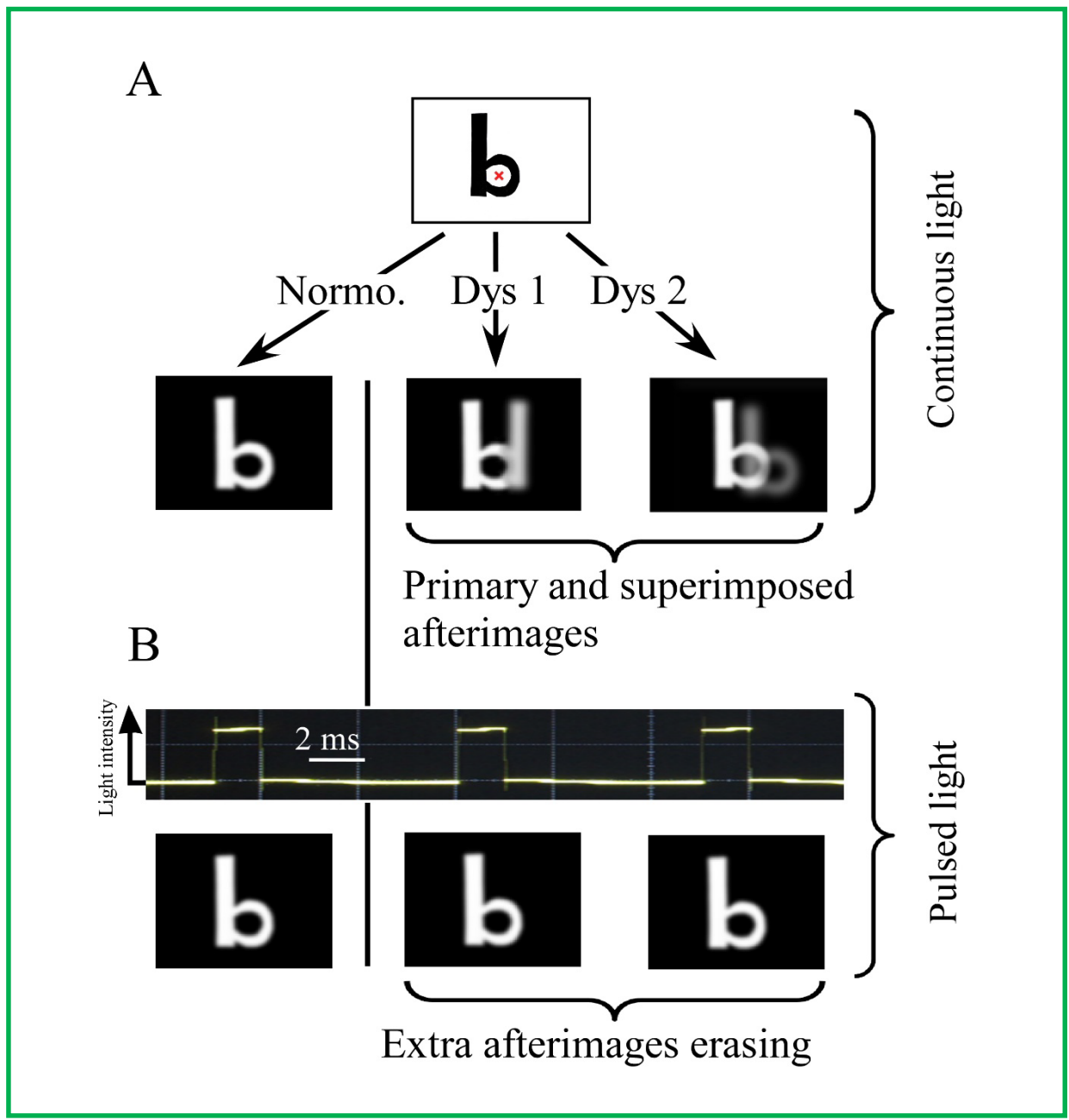

*Figure 4: (A) Afterimages of a 'b' seen by a normal reader and two dyslexic observers in usual continuous lighting. (B) Afterimages seen in pulsed mode around 80 Hz. The extra mirror and duplicated images are erased for the two dyslexic observers. Their afterimages are identical to those of a normal reader.*

Each interhemispheric projection, as shown in Fig. 3C through the corpus callosum, is accompanied by a delay of about 10 ms between the 'b' and the 'd' for example, which corresponds to the time taken for the nerve impulse to travel the ten centimeters or so between the two hemispheres. This delay allows us to try to weaken these too strong projections in the dyslexic person. Using Hebbian processes [22] at the synapses of the cortex, we have proposed pulsed light systems (liquid crystal glasses, lamps, screens, etc…) that can be adjusted electronically to optimise the erasure of extra afterimages. Figure 4 shows the effectiveness of these systems. For a normal reader who has fixed a 'b' as a stimulus, the afterimage is simply a single 'b', whether using conventional continuous lighting (Fig. 4A) or pulsed lighting (Fig. 4B). However, for dyslexics without asymmetry, a mirror 'd' or a second 'b' can be seen simultaneously with the true 'b' (Fig. 4A) in the continuous regime. The use of a pulsed system (here around 80 Hz) leads to the same afterimage as in the normal reader (Fig. 4B). All the distracting projections are erased. The effect is immediate and the method non-invasive. In short, we can have an optical control of the primary cortex. This corresponds to a small part of Crick's hope [23] for separate optical control of each brain region. Moreover, this optical control might also be involved in other functions and other disorders common to dyslexics, such as phonological and magnocellular difficulties, motor skills and posture [24]. As vision is the brain's main source of information from birth, its role in the cross-modal areas is important throughout brain development.

In conclusion, dyslexia might appear in many cases to be a disorder of brain connectivity linked to a lack of asymmetry in Maxwell's centroids and cortical neurons. This lack of asymmetry induces a visual internal crowding which can be compensated by pulsed systems using Hebbian processes at the synapses. New possibilities and new tests may perhaps be suggested by orthoptists and speech therapists, with new early diagnostic tests using the foveascope, making it possible to exploit the great neuronal plasticity in young children.

## Declaration of links of interest

This manuscript is essentially a translation of the paper entitled « Le manque d'asymétrie des centroïdes de Maxwell, et de dominance oculaire, chez les dyslexiques » published in the Revue Francophone d'Orthoptie : Le Floch A, Ropars G. 2020 Le manque d'asymétrie des centroïdes de Maxwell, et de dominance oculaire, chez les dyslexiques. Revue Francophone d'Orthoptie **12**, 134-138. Doi : 10.1016/j.rfo.2020.07.011